# "Different minds" in the jigsaw puzzle of H-bomb story

*Gennady Gorelik*

*https://ggorelik.wordpress.com/*

**Abstract**

The Super-bomb issue contained two problems: feasibility of physical engineering and political need to get H-bomb ASAP. And there were two radically different assessments of these problems by outstanding physicists Hans Bethe and Edward Teller, connected by many years of personal friendship and engaged in the US nuclear project from the very beginning, as the head of the Theoretical Division at Los Alamos and the leading figure in thermonuclear research. This puzzle is solved by using the history of the Soviet H-bomb, in particular the declassified documents on the role of Klaus Fuchs.

*In memory of Freeman Dyson*

The detective puzzle of 6-7 January 1953, presented by Alex Wellerste in "John Wheeler's H-bomb blues" (Physics Today 72, 12, 2019), involved a lot of politics on different levels of government. Since smartest FBI agents failed to solve it on the spot in space-time, it seems to remain unsolvable forever. For physicists of today, however, more interesting might be a physicist puzzle in the core of the story. Seventy years later, this puzzle could be solved, helping to see a likely solution for the detective one.

The Super-bomb issue contained two problems: feasibility of physical engineering and political need to get H-bomb ASAP. And there were two radically different assessments of these problems by outstanding physicists Hans Bethe and Edward Teller, connected by personal friendship since their European 20s and engaged in the US Nuke project from the very beginning, as the head of the Theoretical Division at Los Alamos and the leading figure in thermonuclear research.

In May 1952 (after the first Soviet test of A-bomb in Aug. 1949, and arrest of Klaus Fuchs in Feb. 1950) Hans Bethe wrote in his assessment:

*"In 1951 Teller discovered an entirely new approach to thermonuclear reactions. I believe that among all scientists in the United States he was the only one who could have made this discovery due to his ingenuity and his persistent belief in thermonuclear reactions... Even*

*with Teller, the discovery was largely accidental. The new approach … was based on two separate discoveries, (a) that high densities would be useful and (b) that they could be achieved by a radiation implosion. …Whether the same accidental discoveries have been made in Russia, it is entirely impossible to judge."*

Bethe had no doubts that "*the Russians were very much helped in the development of their fission bomb by information given then by Dr. Klaus Fuchs*" and shared the concern that "*the Russians may be engaged in a major effort to develop the H-bomb*". But he did not believe that "*our information on thermonuclear bombs as of 1946*" (when Fuchs left the US) "*would lead [the Russians] in a rather straightforward way to a successful hydrogen bom*b", - because "*the H-bomb designs for which we now expect success are almost exactly the opposite of those proposed in 1946*". So, "*if the Russians have followed the 1946 line of development, we can only be happy because they would have wasted a lot of effort on a project without military significance*". And Bethe concluded: "*Clearly, no amount of work can assure us of a lasting monopoly in this field. On the contrary, if we now publicly intensify our efforts we shall force the Russians even more into developing this weapon which we have every reason to dread."*

In August 1952 Edward Teller, in his assessment, responded Bethe:

*"The main principle of radiation implosion was developed in connection with the thermonuclear program and was stated at a conference on the thermonuclear bomb, in the spring of 1946. Dr. Bethe did not attend this conference but Dr. Fuchs did. It is difficult to argue to what extent an invention is accidental: most difficult for someone who did not make the invention himself. It appears to me that the* [invention of 1951] *was a relatively slight modification of ideas generally known in 1946. Essentially only two elements had to be added: to implode a bigger volume, and, to achieve greater compression by keeping the imploded material cool as long as possible".* And Teller concluded: *"We may, therefore, be at the beginning of an arduous program and it is quite possible that the Russians have advanced much farther along that road than we have."*

Both opposing assessments remained classified for about 40 years. But just in two years the assessors repeated them in public at the 1954 hearing "In the Matter of J. Robert Oppenheimer".

Bethe*: "there was a very brilliant discovery made by Dr. Teller. It was one of the discoveries for which you cannot plan, one of the discoveries like the discovery of the relativity* *theory, although I don't want to compare the two in importance. But something which is a stroke of genius, which does not occur in the normal development of ideas. But somebody has to suddenly have an inspiration. It was such an inspiration which Dr. Teller had which put the program on a sound basis.* …

*Dr. Teller has a mind very different from mine. I think one needs both kinds of minds to make a successful project. I think Dr. Teller's mind runs particularly to making brilliant inventions, but what he needs is some control, some other person who is more able to find out just what it is the scientific fact about the matter. Some other person who weeds out the bad from the good ideas."*[1]

Teller:  *"I think that it was neither a great achievement nor a brilliant one. It just had to be done. I must say it was not completely easy. There were some pitfalls. But I do believe that if the original plan in Los Alamos, namely, that the laboratory with such excellent people like Fermi and Bethe and others, would have gone after the problem, probably some of these people would have had either the same brilliant idea or another one much sooner. In that case I think we would have had the bomb in 1947. I do not believe that it was a particularly difficult thing as scientific discoveries go. I do not think that we should now feel that we have a safety as compared to the Russians, and think it was just necessary that somebody should be looking and looking with some intensity and some conviction that there is also something there.*"[2]

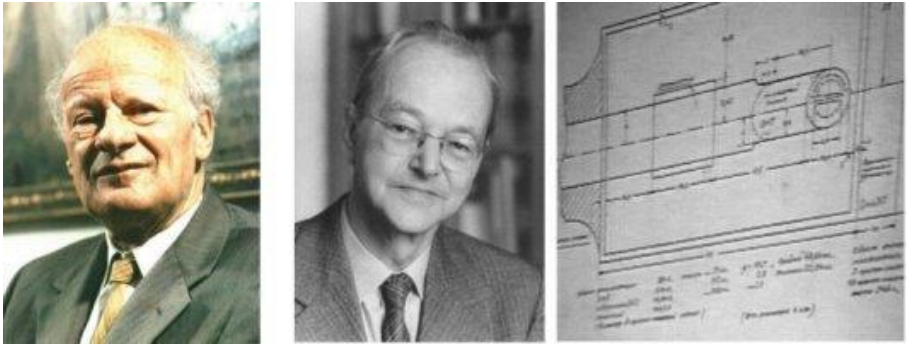

Hans Bethe, Klaus Fuchs, and a sketch of a design for the H-bomb in Fuchs's 1948 intelligence report

The "Oppenheimer Affair" of 1954, being a result of "Wheeler's H-bomb blues" against background of the Red Scare, was summarized by Kip S. Thorne:

---

[1] Reports to the U.S. Atomic Energy Commission. U.S. Government Printing Office, May 1953, p.330, 331.
[2] Reports to the U.S. Atomic Energy Commission. U.S. Government Printing Office, May 1953, p.714.

*"To most American physicists, Oppenheimer became an instant martyr and Teller an instant villain. Teller would be ostracized by the physics community for the rest of his life. But to Wheeler, it was Teller who was the martyr: Teller had "had the courage to express his honest judgment, putting his country's security ahead of solidarity of the community of physicists," Wheeler believed. Such testimony, in Wheeler's view, "deserved consideration," not ostracism. Andrei Sakharov, thirty-five years later, came to agree. Just for the record, I strongly disagree with Wheeler (though he is one of my closest friends and my mentor) and with Sakharov.* " [3]

Freeman Dyson did not belong to majority, and for him "*the main question was whether the security rules should be applied impartially to famous people and unknown people alike. It was a question of fairness. If any unknown person had behaved as Oppenheimer behaved, telling a lie to a security officer about an incident that involved possible spying, he would certainly have been denied clearance. The question was whether Oppenheimer, because he was famous, should be treated differently. Should there be different rules for peasants and princes? This was a question concerning which reasonable people could disagree. I tended to agree with Teller that the rules ought to be impartial. And I saw no reason why people who disagreed with him should condemn him for speaking his mind.* " [4]

Responding to the view of majority, expressed in A. Lightman's review of Teller's memoirs, Dyson wrote in 2005:  "*Lightman admits that there were two Tellers. He writes, "There is a warm, vulnerable, honestly conflicted, idealistic Teller, and there is a maniacal, dangerous, and devious Teller." But his portrait of Teller shows us mostly the dark side. I knew Teller well and worked with him joyfully for three months on the design of a safe nuclear reactor. The Teller that I knew was the warm, idealistic Teller. We disagreed fiercely about almost everything and remained friends. He was the best scientific collaborator I ever had. I consider Lightman's portrayal of him to be unjust. My own review of Teller's memoirs explains why*". [5]

Bethe also did not belong to majority. He opposed Teller in more than one issue, but never questioned his honesty and never changed his opinion about Teller's role in the invention of H-bomb in 1951. The latest indication is Bethe's remark (in biography of J. R. Oppenheimer,

---

[3] Kip Thorne. Black Holes and Time Warps: Einstein's Outrageous Legacy. W. W. Norton, 1994, p.235.
[4] Freeman J Dyson. Birds and Frogs: Selected Papers, 1990-2014. World Scientific, 2015, p. 134.
[5] F. Dyson, Seeing the unseen. The New York Review of Books, February, 2005.

1997): "*the crucial invention was made in 1951, by Teller*".[6] Bethe never used expression like widely accepted "Teller-Ulam design" though stressed mathematical contribution of Stanislaw Ulam who in 1950 "on his own initiative"[7] checked (un)feasibility of the initial design of H-bomb (Classical Super).

To explain the disagreement in assessments of most prominent and mutually respected experts, a philosophical key could be found in Bethe's words: "*Dr. Teller has a mind very different from mine. I think one needs both kinds of minds to make a successful project.*" Any assessment relies on inevitably limited knowledge of the assessor and his inevitably subjective (and mysterious) intuition. Great minds in physics could be very different indeed. Since Bethe mentioned the relativity theory, I would remind that Einstein praised "*Bohr's unique instinct and tact to discover the major laws of the spectral lines*" as "*a miracle*" and actually acknowledged his own inability to make such an invention.

Philosophy, however, is too general and uncertain tool to solve such a specific jigsaw puzzle as the history of American H-bomb. Having in mind that this history was not confined within the US borders, the factual tool could be found in biography of the Soviet counterpart of Teller - Andrei Sakharov, who three years after Teller made the same "*brilliant discovery*", in 1958 opposed Teller in the issue of "clean H-bomb," in 1975 was awarded the Nobel Peace Prize for his human rights activity, in the 1980's opposed Teller in the issue of SDI, and in the 1989 wrote:

"*There are very strong arguments in favor of Teller's point of view* [in the late 1940s and early1950s], *based on our knowledge of the real situation in the world at that time. The government of the USSR, or rather those who were in power - Stalin, Beria and others, already knew about the potential of the new weapon and in no case would not abandon efforts to create it. Any American steps to temporarily or permanently stop the development of thermonuclear weapons would be regarded either as a cunning, deceptive, distracting maneuver, or as a manifestation of stupidity or weakness. In both cases, the reaction would be unambiguous - not to fall into the trap, and immediately use the stupidity of the enemy. ... As to the attitude of American colleagues to Teller, it seems to me unfair (and even mean-spirited). Teller had taken*

[6] H. Bethe. The Road from Los Alamos. Springer, 1991, p.228.
[7] H. Bethe. Comments on the History of the H-Bomb. Los Alamos Science, 1980, Vol. 1, N 1, p. 42.

*a stand based on principle. The very fact that Teller went against the current, against the majority opinion, speaks in his favor.*" [8]

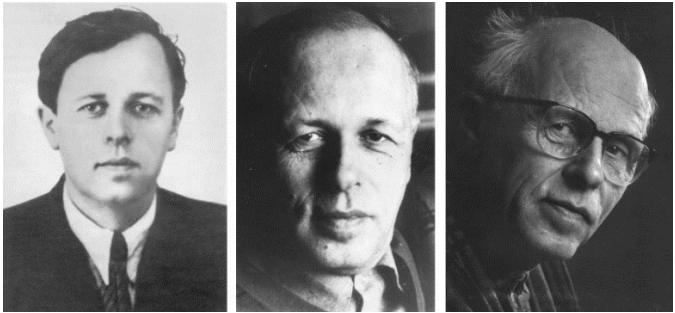

Andrei Sakharov in 1950, 1980 and 1989

For half a century the jigsaw puzzle of H-bomb story was kept in two secret boxes separated by the Iron Curtain of Cold War. Massive declassification of the Soviet nuke history started in 1995 four years after the collapse of the Soviet system and six years after Sakharov's death. It resulted in the multi-volume series "The Soviet atomic project: documents and materials" [9] including two volumes on H-bomb.

Sakharov didn't know about some of the most important documents in this collection, but he knew first-hand that the Soviet "crash program" for H-bomb started in June 1948 when he was drafted together with a few of his colleagues and their mentor Igor Tamm at the academic Physical Institute (FIAN). This auxiliary group was established to help with calculations for the Yakov Zeldovich's group, which since 1946 had been working on the H-bomb design dubbed "Truba" ("Tube", for its cylindrical shape). In two months, Sakharov's intuition hinted that the design was not promising, and he invented an entirely different alternative design that employed a special way of compressing a spherically layered configuration, dubbed "Sloyka" (Layercake). In the fall of 1948 the Soviet nuclear establishment endorsed the new design, and Tamm's group concentrated on it. Thus, well before the start of American "crash program" on H-bomb, in the USSR they started work on two designs of H-bomb.

Sakharov didn't know that Zeldovich's Tube was "imported" from the US Classical Super, that it was detailed intelligence report by Klaus Fuchs of April 1948 urged Stalin to boost H-bomb effort, that this report contained Fuchs' 1946 idea of compression by radiation, and that Zeldovich failed to see the potential of this idea and had ignored it.

---

[8] А. Сахаров. Воспоминания. М.: Права человека, 1996, Т. 1, *с. 144-146*.
[9] Атомный проект СССР: документы и материалы (1998-2010). http://elib.biblioatom.ru/sections/0201

It was good luck for humanity and for Sakharov that Beria, the head of the Soviet nuclear project, in March 1949 declined the request of Yuli Khariton, the scientific head of the Soviet nuke program, to permit Tamm access to the intelligence data – "*in order not to attract unnecessary persons to these documents*." [10] By that time Sakharov was prepared to grasp Fuchs' idea much better than Teller, because Sakharov's Layercake was based on compression and in the report on Layercake, he proposed the "*use of an additional plutonium charge for a preliminary compression of Layercake*". [11] It was two years before Ulam's idea of "two bombs in a box" which is considered by many as the reason to use the term "Teller-Ulam design". And fusion fuel Li6 was proposed by Vitaly Ginzburg in the fall of 1948, two years earlier than in the US.

After the arrest of Fuchs the simplest secret of H-bomb was that Classical Super (aka Tube) was a dead end. In the US it was discovered in early 1950, thanks to Ulam, while it the USSR as late as the end of 1953. Only after the closure of the initial projects the successful super-bomb designs were invented, thanks to creative intuitions of Teller and Sakharov, independently in the US and the USSR. No wonder that Zeldovich spoke about Sakharov (like Bethe about Teller): "*I can understand and take the measure of other physicists, but Andrei* [Sakharov] *-- he's something else, something special.*" [12]

Teller's belief that the H-bomb was attainable not only by his "accidental stroke of genius" defined his political assessment. When Sakharov - in hindsight - endorsed Teller's assessment, he relied on his personal experience of a top Soviet expert in strategic weaponry.

Teller "declassified" his personal source of information only after the collapse of the Soviet system:

"*The events in the Soviet Union got an emotional emphasis when my good friend, the excellent physicist, Lev Landau, was jailed* [in 1938]. *I had known him in Leipzig and Copenhagen as an ardent Communist. I was pushed to the conclusion that Stalin's Communism was not much better than the Nazi dictatorship of Hitler.*" [13]

---

[10] http://elib.biblioatom.ru/text/atomny-proekt-sssr_t3_kn1_2008/go,186/?bookhl=%22чтобы+не+привлекать+к+этим+документам+лишних+людей%22

[11] http://elib.biblioatom.ru/text/atomny-proekt-sssr_t3_kn1_2008/go,168/?bookhl=%22использование+дополнительного+заряда+плутония+для+предварительного+сжатия+слойки%22

[12] G. Gorelik. The World of Andrei Sakharov: A Russian Physicist's Path to Freedom (transl. A. W. Bouis). Oxford University Press, 2005, p. 188.

[13] G. Gorelik. The World of Andrei Sakharov, p. 212.

*"My second published paper in physics was a joint undertaking with my good Hungarian friend, Laszlo Tisza. Shortly after our collaboration in Leipzig he was arrested as a communist by the Hungarian fascist government. He had lost his chance of obtaining an academic position and I referred him, with my strong recommendation, to my friend Lev Landau in Kharkov, Ukraine. A few years later Tisza visited me in the United States. He no longer had any sympathy with Communism. Lev Landau had been arrested in the Soviet Union as a capitalist spy! The implication of this event was for me even more defining than the Hitler-Stalin Pact. By 1940, I had every reason to dislike and distrust the Soviets."* [14]

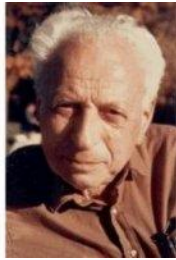

Laszlo Tisza (1907-2009), #5 in the list of Lev Landau's graduates

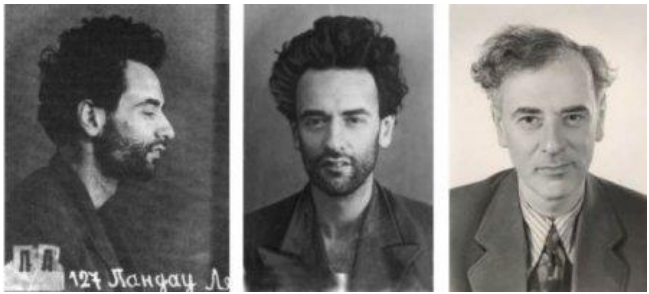

Lev Landau in prison, 1938, and twenty years later

For Teller the reasons to build the H-bomb to counter Stalin's Communism were no less serious than to build the A-bomb to counter Hitler's Nazism. In 2001 Teller was surprised and pleased to learn that Landau in 1957 (courtesy of the KGB wiretapping) spoke:

*"Our regime, as I know it from 1937 on, is definitely a fascist regime, and it could not change by itself in any simple way . . . Our leaders are fascists from head to toe. They can be more liberal or less liberal, but their ideas are fascist."* [15]

---

[14] E. Teller. Science and Morality. Science 1998, Vol. 280, Issue 5367, p. 1200. Cf. L. Tisza, Adventures of a Theoretical Physicist. Physics in Perspective 11, 46–97 (2009).

[15] G. Gorelik. Lev Landau, Prosocialist Prisoner of the Soviet State. Physics Today 48, 11 (1995); The top-secret life of Lev Landau. Scientific American 1997, 277 (2), 72-77.

Writing his Memoirs in the 1980s, Sakharov reflected on how he could think that he was committed to the same goal as Stalin did – "*building up the nation's strength to ensure peace after a devastating war*" and concluded with self-assessment: "*I was unwittingly… creating an illusory world to justify myself*". [16]

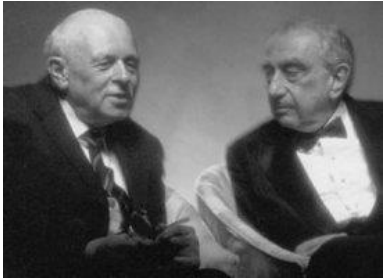

Andrei Sakharov and Edward Teller, 1988

All assessments are, in a sense, illusory worlds. The difference is how close to reality is this or that illusory world. The history of the American and Soviet H-bombs helps to see that Teller's assessment was closer to reality than assessments by many his colleagues.

A real physicist has to understand the role of empirical evidence in personal experience, including very special experience of Klaus Fuchs, about which Teller wrote in 1950 to his colleague and friend Maria Goeppert-Mayer:

*"You remember Klaus Fuchs? …He was too reserved for my taste although he was always very nice. He must have been living under an incredible stress. Quite a few people here [at Los Alamos] are furious at Fuchs. They feel personally insulted. I do not feel that way. We should have learned what kind of a system the communist party is and what kind of demands it makes on its members. Fuchs probably decided when he was 20 years old (and when he saw Nazism coming in Germany) that the communists are the only hope. He decided that before he ever became a scientist. From that time on his whole life was built around that idea.*

*People always do this: They underestimated the Nazis and they underestimate now the communists. Then the disaster comes, and then the same people who would not believe that trouble is ahead get very angry at individual communists or individual Nazis.*

*Actually, the damage that Fuchs has done is great. He surely gave away a lot, and by now I feel quite doubtful whether we can keep up with the Russians in the atomic race. But that is not all. The Fuchs case probably will give rise to great difficulties in making a sensible*

---

[16] G. Gorelik. The World of Andrei Sakharov, p. 165.

*agreement with the British (and it seemed just around the corner). And finally, we can now confidently expect a witch-hunt.*" [17]

In the 1950s  personal experience of many American physicists made much  easier to dislike capitalist government of the US than to understand the real life of "scientific socialism" behind Iron Curtain totally controlled by the Soviet autocracy.

As to the secret document vanished from Wheeler's envelope in 1953, the history of the Soviet H-bomb just has no place for this document.  It's easy to imagine a person peeking into a stranger's envelope (e.g. to look for the name of the owner) and seeing the terrifying stamp "Top Secret". This person most probably would have been stung by the realization that his fingerprints on this damned paper would lead him to the same electric chair that awaited the Rosenbergs, and felt he had to annihilate it immediately  (e.g. by eating).

More detailed arguments and all the references are in the article[18].

---

P. S.

**Subject:**     Re: Physics Today Support - Submit: Letter to Editor - Ticket # 1672
**Date:**         Tue, 03 Mar 2020 14:16:14 +0000

Dear Dr. Gennady Gorelik,

We appreciate receiving the article proposal that you sent to *Physics Today*, titled "Different Minds in the Jigsaw Puzzle of H-bomb Story...."  A committee of our editors recently met to discuss several article proposals that we received, including yours.  The committee determined that it does not meet our editorial needs at this time.

As a magazine that serves the scientific community, we strive to publish as many responsible voices as we can.  Due to the volume of submissions we receive, however, we must decline a large number of submissions.

Thank you for your understanding and patience, and for your interest in *Physics Today*.

Sincerely,

pteditors@aip.org

---

[17] E. Teller. Memoirs: A Twentieth Century Journey in Science and Politics. Basic Books, 2009, p. 275-276.
[18] G. Gorelik. The Paternity of the H-Bombs: Soviet-American Perspectives. Physics in Perspective 2009, 11(2), 169-197.